\documentclass[twocolumn,letter]{jpsj3}
\usepackage{txfonts}
\usepackage{graphicx}
\usepackage{dcolumn}
\usepackage{bm}
\usepackage[usenames,dvipsnames]{xcolor}
\usepackage{ulem}
\usepackage{braket}
\usepackage{amsmath}
\usepackage{tabularx}

\title{Degeneracy Lifting of N\'eel, Bloch, and Anti-Skyrmion Crystals in Centrosymmetric Tetragonal Systems}

\author{Satoru Hayami$^1$ and Ryota Yambe$^2$}
\inst{
$^1$Department of Applied Physics, University of Tokyo, Bunkyo, Tokyo 113-8656, Japan \\
$^2$Department of Physics, Hokkaido University, Sapporo 060-0810, Japan \\
 }

\abst{
A skyrmion crystal has internal degrees of freedom in terms of vorticity and helicity, which gives rise to fascinating types of skyrmion phases (N\'eel, Bloch, and anti-skyrmion).  
We show how the degeneracy lifting in tetragonal skyrmion crystals occurs in centrosymmetric itinerant magnets by focusing on microscopic interactions through the Fermi surface instability. 
We demonstrate that four types of skyrmion crystals can be realized according to the signs of the anisotropic Ruderman-Kittel-Kasuya-Yosida interaction, which are sensitive to band structures, electron fillings, atomic orbitals, and sublattice positions. 
Our findings might provide a possibility of controlling the types of the skyrmion crystals by electron doping and external pressure rather than relying on the noncentrosymmetric lattice structures. 
}

\begin{document}
\maketitle

A magnetic skyrmion, whose magnetic texture is characterized by a topological number~\cite{skyrme1962unified,Bogdanov89,Bogdanov94,rossler2006spontaneous,nagaosa2013topological}, has attracted much interest in condensed matter physics since the experimental identification of the magnetic skyrmion crystals (SkX) by neutron scattering~\cite{Muhlbauer_2009skyrmion} and Lorentz transmission electron microscopy~\cite{yu2010real}. 
The SkXs are classified into their vorticity $l$ and helicity $\gamma$, such as the Bloch SkX~\cite{Muhlbauer_2009skyrmion,yu2010real,seki2012observation,tokunaga_new_2015,karube2016robust}, N\'eel SkX~\cite{kezsmarki_neel-type_2015,Kurumaji_PhysRevLett.119.237201}, anti-SkX~\cite{koshibae2016theory,nayak2017discovery,hoffmann2017antiskyrmions,saha2019intrinsic}, and bi-SkX~\cite{yu2014biskyrmion,lee2016synthesizing}.  
As the different types of the SkXs give rise to different magneto-electric responses~\cite{Gobel_PhysRevB.99.060406} and dynamics~\cite{zhang2017skyrmion,Rozsa_PhysRevB.95.094423,kovalev2018skyrmions,liang2018magnetic,McKeever_PhysRevB.99.054430,peng2020controlled}, their control might be promising for potential applications to next-generation spintronic devices. 

\begin{figure*}[htb!]
\begin{center}
\includegraphics[width=1.0 \hsize]{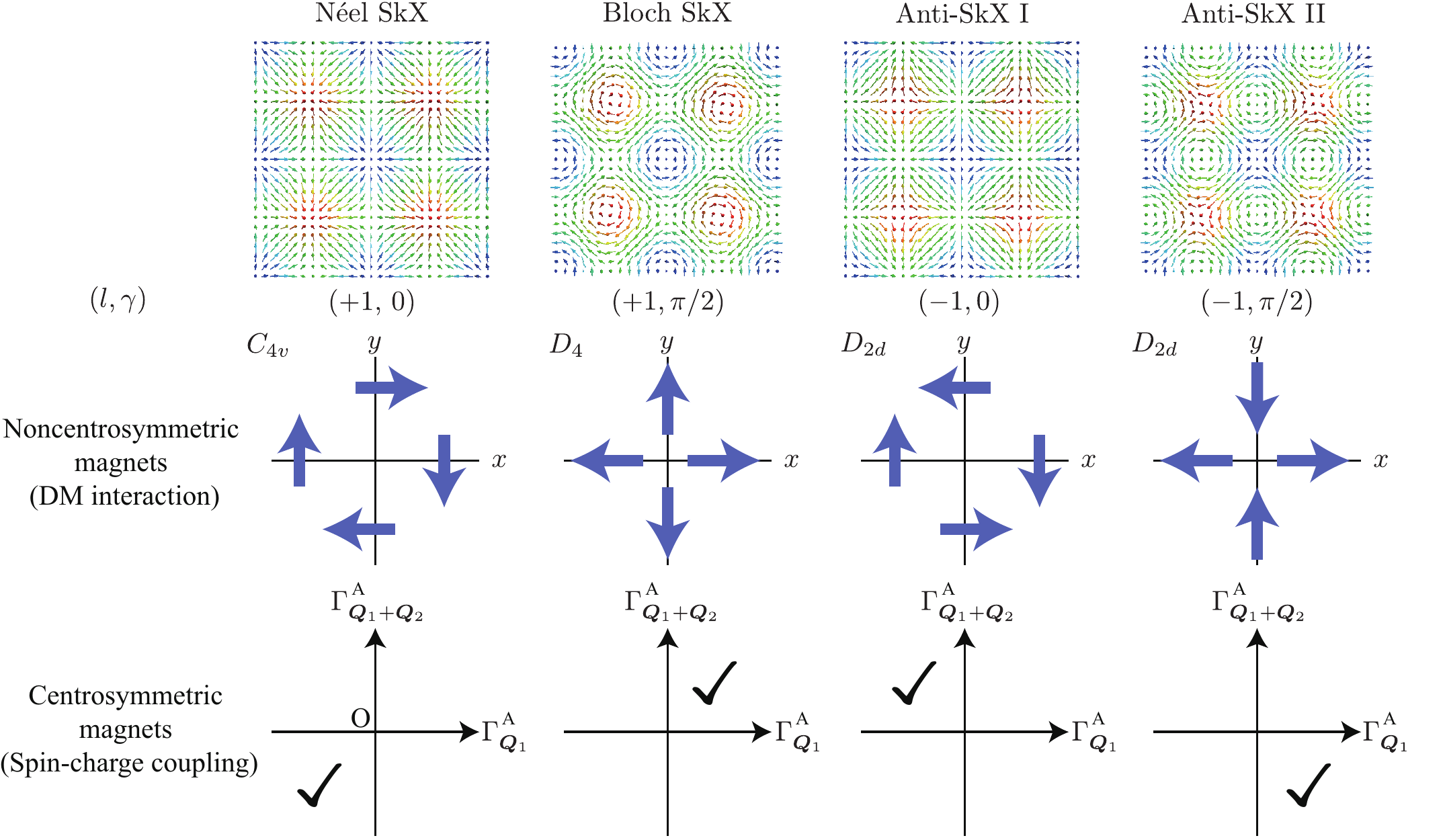} 
\caption{
\label{Fig:ponti}
(Color online) (Upper panel) Classification of the skyrmion crystals (SkXs) according to their vorticity $l$ and helicity $\gamma$: the N\'eel SkX with $(l, \gamma)=(+1,0)$, the Bloch SkX with $(+1, \pi/2)$, the anti-SkX I with $(-1,0)$, and the anti-SkX II with $(-1, \pi/2)$. 
The middle panel shows the case of the SkXs in the presence of the DM interaction in noncentrosymmetric magnets.  
The blue arrows represent the DM vector on the nearest-neighbor bonds on a square lattice where the corresponding point groups are shown. 
The bottom panel shows the signs of the effective anisotropic interactions to lift the degeneracies of $(l, \gamma)$ through the spin-charge coupling in centrosymmetric itinerant magnets. 
}
\end{center}
\end{figure*}

In noncentrosymmetric magnets, it has been well-known that the SkXs are described~\cite{nagaosa2013topological,Banerjee_PhysRevX.4.031045,leonov2016properties} by the spin model with the ferromagnetic interaction and the Dzyaloshinskii-Moriya (DM) interaction, the latter of which originates from the spin-orbit coupling~\cite{dzyaloshinsky1958thermodynamic,moriya1960anisotropic}. 
In particular, the types of the SkXs are closely related to the direction of the DM vector determined from the point group symmetry. 
When the DM vector lies in the direction perpendicular to the bond, the N\'eel SkX with $l=1$ and $\gamma=0, \pi$ or the anti-SkX (refereed as anti-SkX I) with $l=-1$ and $\gamma=0, \pi$ is stabilized, whereas the Bloch SkX with $l=1$ and $\gamma=\pm \pi/2$ or another anti-SkX (referred as anti-SkX II) with $l=-1$ and $\gamma=\pm \pi/2$ is stabilized when the DM vector is parallel to the bond direction. 
The difference between the N\'eel (Bloch) SkX and two anti-SkXs is found in the vorticity, which is attributed to the presence of the fourfold rotational or fourfold rotational-inversion symmetry in crystal systems. 
Thus, one can naturally expect the types of the SkXs 
in noncentrosymmetric materials according to the point group symmetry~\cite{Gungordu_PhysRevB.93.064428}. 
For example, the N\'eel, Bloch, and anti-SkXs emerge under the point groups $C_{4v}$, $D_{4}$, and $D_{2d}$, respectively, in the case of the tetragonal crystal systems, as shown in Fig.~\ref{Fig:ponti}. 

In contrast to noncentrosymmetric magnets, the SkXs in centrosymmetric magnets have degenerate internal degrees of freedom in terms of $l$ and $\gamma$~\cite{Okubo_PhysRevLett.108.017206,leonov2015multiply,Lin_PhysRevB.93.064430,Hayami_PhysRevB.93.184413,batista2016frustration}. 
This is because dominant interactions to stabilize the SkXs are a competing exchange interaction in the frustrated magnets~\cite{Okubo_PhysRevLett.108.017206,leonov2015multiply,Lin_PhysRevB.93.064430,Hayami_PhysRevB.93.184413,batista2016frustration} and a biquadratic interaction through the spin-charge coupling in itinerant magnets~\cite{Martin_PhysRevLett.101.156402,Akagi_PhysRevLett.108.096401,Hayami_PhysRevB.90.060402,Ozawa_doi:10.7566/JPSJ.85.103703,Hayami_PhysRevB.95.224424,Ozawa_PhysRevLett.118.147205}, which do not select specific $l$ and $\gamma$.  
Meanwhile, in real materials, such a degeneracy should be lifted owing to the presence of additional small interactions, such as the crystalline magnetic anisotropy and dipole-dipole interactions. 
Moreover, the spin-orbit coupling might play an important role in lifting the degeneracy even in centrosymmetric magnets. 
In fact, the recent observations in centrosymmetric magnets indicate the definite type of the SkXs in various lattice systems, such as the triangular-lattice system Gd$_2$PdSi$_3$~\cite{kurumaji2019skyrmion,Hirschberger_PhysRevLett.125.076602,Hirschberger_PhysRevB.101.220401}, the trimer system Gd$_3$Ru$_4$Al$_{12}$~\cite{hirschberger2019skyrmion}, and the square-lattice system GdRu$_2$Si$_2$~\cite{khanh2020nanometric}.  

In the present study, we demonstrate how the degeneracy in terms of $l$ and $\gamma$ is lifted in the SkXs in centrosymmetric lattice structures.
By considering an effective model with anisotropic bilinear-biquadratic interactions from a symmetry viewpoint, we find that anisotropic Ruderman-Kittel-Kasuya-Yosida (RKKY) interactions depending on the wave vectors play an important role in determining the definite $l$ and $\gamma$. 
We realize four types of the SkXs (N\'eel, Bloch, and two anti-SkXs) on a square lattice by using the simulated annealing. 
We clarify that there are two key anisotropic parameters to determine the types of the SkXs in the centrosymmetric tetragonal point group $D_{4h}$: one is the sign of the anisotropic interaction in the ordering-vector channel, and the other is the sign of the anisotropic interaction in the higher harmonic channel consisting of the addition of the two-ordering vectors.  
Our results indicate that centrosymmetric itinerant magnets will provide a fertile playground for searching for further exotic SkXs through the electron correlation. 

Let us start by classifying the SkXs from the symmetry viewpoint. 
We here consider the four types of the SkXs where each skyrmion core periodically aligns in a square-lattice way under the strong fourfold rotational anisotropy, as shown in Fig.~\ref{Fig:ponti}. 
Since all the SkXs break both the spatial inversion and time-reversal symmetries, they are classified into the rotational and mirror symmetries. 
As the vorticity $l$ stands for the winding number of the spin textures projected onto the $xy$ spin plane, it is related to the rotational symmetry around the $z$ axis, namely, $l=+1$ in the N\'eel and Bloch SkXs corresponds to the fourfold rotational symmetry, while $l=-1$ in the two anti-SkXs corresponds to the fourfold rotational-inversion symmetry. 
On the other hand, the helicity $\gamma$ is related to the [100] mirror symmetry regarding the $xy$ spin component, i.e., absence (presence) of the [100] mirror plane indicates $\gamma=0, \pi$ ($\gamma=\pm \pi/2$) in the N\'eel (Bloch) SkX. 

The square-lattice SkXs are characterized by a superposition of two cycloidal or proper-screw spirals, whose spin textures are given by 
\begin{align}
\label{eq:SkX}
\bm{S}_i =  
\left(
\begin{array}{c}
c_x \sin (\bm{Q}_1 \cdot \bm{r}_i + \theta_1)+c'_x \sin (\bm{Q}_2 \cdot \bm{r}_i + \theta'_2)\\ 
c_y \sin (\bm{Q}_2 \cdot \bm{r}_i + \theta_2)+c'_y \sin (\bm{Q}_1 \cdot \bm{r}_i + \theta'_1)\\
\tilde{M}-c_z \left[\cos (\bm{Q}_1 \cdot \bm{r}_i + \theta''_1)+\cos (\bm{Q}_2 \cdot \bm{r}_i + \theta''_2) \right]
\end{array}
\right), 
\end{align}
where $c_x, c_y, c_z, c'_x, c'_y$, $\tilde{M}$, $\theta_\eta$, $\theta'_\eta$, and $\theta''_\eta$ ($\eta=1,2$) are numerical coefficients depending on the model parameters. 
When the ordering vectors $\bm{Q}_1$ and $\bm{Q}_2$ are oriented along the $x$ and $y$ directions, respectively, the four types of the SkXs in Fig.~\ref{Fig:ponti} are represented as follows: $c_x=c_y$ and $c'_x=c'_y=0$ in the N\'eel SkX, $c_x=c_y=0$ and $-c'_x=c'_y$ in the Bloch SkX, $c_x=-c_y$ and $c'_x=c'_y=0$ in the anti-SkX I, and $c_x=c_y=0$ and $c'_x=c'_y$ in the anti-SkX II.

One of the four-types of the SkXs can be unambiguously selected in noncentrosymmetric crystal structures owing to the emergent DM interaction.  
In other words, the types of the SkXs are expected from the noncentrosymmetric point groups: the N\'eel SkX is induced under the point group $C_{4v}$, the Bloch SkX is induced under the point group $D_{4}$, and the two anti-SkXs are induced under the point group $D_{2d}$. 
It is noted that the helicity (0 or $\pi$ in the N\'eel SkX and anti-SkX I, and $+\pi/2$ or $-\pi/2$ in the Bloch SkX and anti-SkX II) is determined by the sign of the DM vector. 
The correspondence among the SkX spin textures, DM interactions, and the point groups is shown in Fig.~\ref{Fig:ponti}. 

Meanwhile, in centrosymmetric magnets where the DM interaction is absent, the exchange interactions through the electron correlation play an important role in determining $l$ and $\gamma$. 
In this case, symmetric anisotropic interactions narrow down the types of the SkXs. 

To demonstrate that, we consider an effective spin model with the bilinear and biquadratic interactions in momentum space, which is obtained from the Kondo lattice model consisting of itinerant electrons and localized spins~\cite{Hayami_PhysRevB.95.224424,Hayami_PhysRevLett.121.137202,hayami2020multiple,Su_PhysRevResearch.2.013160,Yasui2020}. 
The model Hamiltonian is given by
\begin{align}
\label{eq:Model}
\mathcal{H}=  &2\sum_\nu
\left[ -J  \left(\sum_{\alpha,\beta}\Gamma^{\alpha\beta}_\mathbf{Q_{\nu}} S^\alpha_{\mathbf{Q_{\nu}}} S^\beta_{-\mathbf{Q_{\nu}}}\right)
+\frac{K}{N} \left(\sum_{\alpha,\beta}\Gamma^{\alpha\beta}_\mathbf{Q_{\nu}} S^\alpha_{\mathbf{Q_{\nu}}} S^\beta_{-\mathbf{Q_{\nu}}}\right)^2 \right]  \nonumber \\
&-H \sum_i S_i^z,
\end{align}
where $N$ is the system size and $\alpha, \beta=x,y,z$. 
$\bm{S}_{\bm{Q}_\eta}$ ($\eta=1,2$) is the Fourier transform of the localized spin $\bm{S}_i$, which is regarded as a classical vector with $|\bm{S}_i|=	1$. 
We take the ordering vectors $\bm{Q}_1=(\pi/3,0)$ and $\bm{Q}_2=(0,\pi/3)$, which are set by the nesting of the Fermi surfaces. 
The following analysis is applied to the other ordering vectors whenever the magnetic susceptibility of itinerant electrons shows maximum peaks at their positions. 
The first term in Eq.~(\ref{eq:Model}) represents the RKKY interaction with the coupling constant $J>0$, while the second term represents the biquadratic interaction with $K>0$. 
$\Gamma^{\alpha\beta}_{\bm{Q}_\eta}$ is a form factor to represent the magnetic anisotropy~\cite{Yasui2020}. 
By assuming the centrosymmetric point group $D_{4h}$, the symmetry analysis leads to the form factors at $\bm{Q}_1$ and $\bm{Q}_2$ represented as  
\begin{align}
\label{eq:Gamma1}
\Gamma_{\bm{Q}_1}=\left(
\begin{array}{ccc}
\Gamma^{xx} & 0 & 0\\
0 & \Gamma^{yy} &0 \\
0 & 0 & \Gamma^{zz}
\end{array}
\right), \ \
\Gamma_{\bm{Q}_2}=\left(
\begin{array}{ccc}
\Gamma^{yy} & 0 & 0\\
0 & \Gamma^{xx} &0 \\
0 & 0 & \Gamma^{zz}
\end{array}
\right).   
\end{align}
We also introduce $\Gamma^{\rm A}_{\bm{Q}_1}=\Gamma^{yy}-\Gamma^{xx}$ to represent the in-plane (bond-dependent) anisotropy. 
It is noted that $\Gamma^{\alpha\beta}_{\bm{Q}_\eta}$ is invariant under the fourfold rotational symmetry. 
The third term stands for the Zeeman coupling to an external magnetic field $H$ along the $z$ direction. 
We here do not consider a crystalline anisotropy in the form of the single-ion anisotropy, since it does not contribute to the degeneracy lifting in SkXs.

\begin{figure}[t!]
\begin{center}
\includegraphics[width=1.0 \hsize]{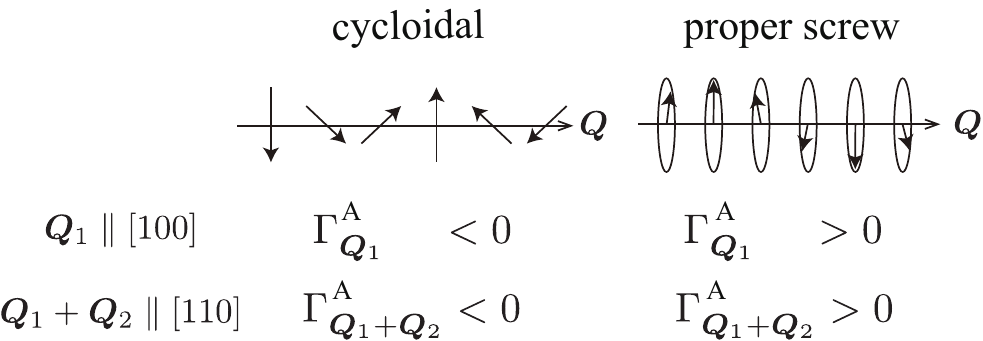} 
\caption{
\label{Fig:int}
The cycloidal and proper-screw spirals in the directions of $\bm{Q}_1$ and $\bm{Q}_1+\bm{Q}_2$ depending on the sign of the anisotropic interactions $\Gamma^{\rm A}_{\bm{Q}_1}$ and $\Gamma^{\rm A}_{\bm{Q}_1+\bm{Q}_2}$. 
}
\end{center}
\end{figure}

The form factors $\Gamma^{\alpha\beta}_{\bm{Q}_\eta}$ determine the type of the spirals; 
For $\Gamma_{\bm{Q}_1}^{\rm A}< 0$ ($\Gamma_{\bm{Q}_1}^{\rm A}> 0$), the cycloidal (proper-screw) spiral state is stabilized in the ground state for $K=H=0$ in Fig.~\ref{Fig:int} (Note that the degeneracy between clockwise and counterclockwise spirals still remains)~\cite{comment_Gammazz}. 
This indicates a possibility to have the N\'eel SkX or anti-SkX I for $\Gamma_{\bm{Q}_1}^{\rm A}< 0$, since they are formed by the superposition of two cycloidal spirals~\cite{Hayami_PhysRevLett.121.137202}. 
In a similar manner, the Bloch SkX or anti-SkX II can appear for $\Gamma_{\bm{Q}_1}^{\rm A}>0$. 
In fact, one of the authors and his collaborators clarify that the Bloch SkX or anti-SkX II is stabilized in the intermediate magnetic field region for nonzero $K$, $H$, and $\Gamma_{\bm{Q}_1}^{\rm A}>0$~\cite{Yasui2020}. 
Note that there is a still degeneracy in terms of $l$, i.e., a degeneracy between N\'eel SkX and anti-SkX I (Bloch SkX and anti-SkX II) in the model in Eq.~(\ref{eq:Model}). 

Under these circumstances, there is a natural question: how is the remaining degeneracy lifted?
To address the question, we focus on the multiple-$Q$ nature in the SkXs.
The SkX spin textures in Eq.~(\ref{eq:SkX}) include the higher-harmonic components of the ordering vectors, such as $\bm{Q}_1+\bm{Q}_2$, which also contribute to the total energy, although their contributions have usually been neglected as small. 
However, as we will describe below, the {\it infinitesimally small} contributions at $\bm{Q}_1+\bm{Q}_2$ and $\bm{Q}_1-\bm{Q}_2$ are enough to lift the remaining degeneracy by combining $\Gamma_{\bm{Q}_1}^{\rm A}$. 

The form factors $\Gamma^{\alpha\beta}_{\bm{Q}_1+\bm{Q}_2}$ and $\Gamma^{\alpha\beta}_{\bm{Q}_1-\bm{Q}_2}$ are given as 
\begin{align}
\label{eq:q1plusq2}
\Gamma_{\bm{Q}_1+\bm{Q}_2}&=\left(
\begin{array}{ccc}
\Gamma'^{xx} & -\Gamma'^{xy} & 0\\
-\Gamma'^{xy} & \Gamma'^{xx} &0 \\
0 & 0 & \Gamma'^{zz}
\end{array}
\right), \nonumber \\
\Gamma_{\bm{Q}_1-\bm{Q}_2}&=\left(
\begin{array}{ccc}
\Gamma'^{xx} & \Gamma'^{xy} & 0\\
\Gamma'^{xy} & \Gamma'^{xx} &0 \\
0 & 0 & \Gamma'^{zz}
\end{array}
\right), 
\end{align}
where we use the notation $\Gamma^{\rm A}_{\bm{Q}_1+\bm{Q}_2}=\Gamma'_{xy}$ hereafter. 
Among the parameters in Eq.~(\ref{eq:q1plusq2}), $\Gamma^{\rm A}_{\bm{Q}_1+\bm{Q}_2}$ lifts a part of the degeneracy in the SkXs in a different way from $\Gamma_{\bm{Q}_1}^{\rm A}$; 
$\Gamma_{\bm{Q}_1+\bm{Q}_2}^{\rm A}< 0$ ($\Gamma_{\bm{Q}_1+\bm{Q}_2}^{\rm A}> 0$) favors the cycloidal (proper-screw) spiral state in Fig.~\ref{Fig:int}~\cite{comment_Gammazz}, which indicates that $\Gamma_{\bm{Q}_1+\bm{Q}_2}^{\rm A}< 0$ ($\Gamma_{\bm{Q}_1+\bm{Q}_2}^{\rm A}> 0$) gives rise to the N\'eel SkX or anti-SkX II (Bloch SkX or anti-SkX I). 

The sign set of $\Gamma^{\rm A}_{\bm{Q}_1}$ and $\Gamma^{\rm A}_{\bm{Q}_1+\bm{Q}_2}$ can lift the degeneracy in terms of $l$ and $\gamma$ in the SkXs, although the individual only lifts a part of the degeneracy.  
This is because the spin textures in the N\'eel SkX and anti-SkX I (Bloch SkX and anti-SkX II), which are degenerate in Eq.~(\ref{eq:Gamma1}), show different oscillations along the $\langle 110 \rangle$ direction. 
Such different spin textures along the $\langle 110 \rangle$ direction are affected by $\Gamma^{\rm A}_{\bm{Q}_1+\bm{Q}_2}$ in Eq.~(\ref{eq:q1plusq2}), which results in lifting the degeneracy. 
We display the sign set of $\Gamma^{\rm A}_{\bm{Q}_1}$ and $\Gamma^{\rm A}_{\bm{Q}_1+\bm{Q}_2}$ in each SkX in the lower panel in Fig.~\ref{Fig:ponti}.

\begin{figure}[t!]
\begin{center}
\includegraphics[width=0.75 \hsize]{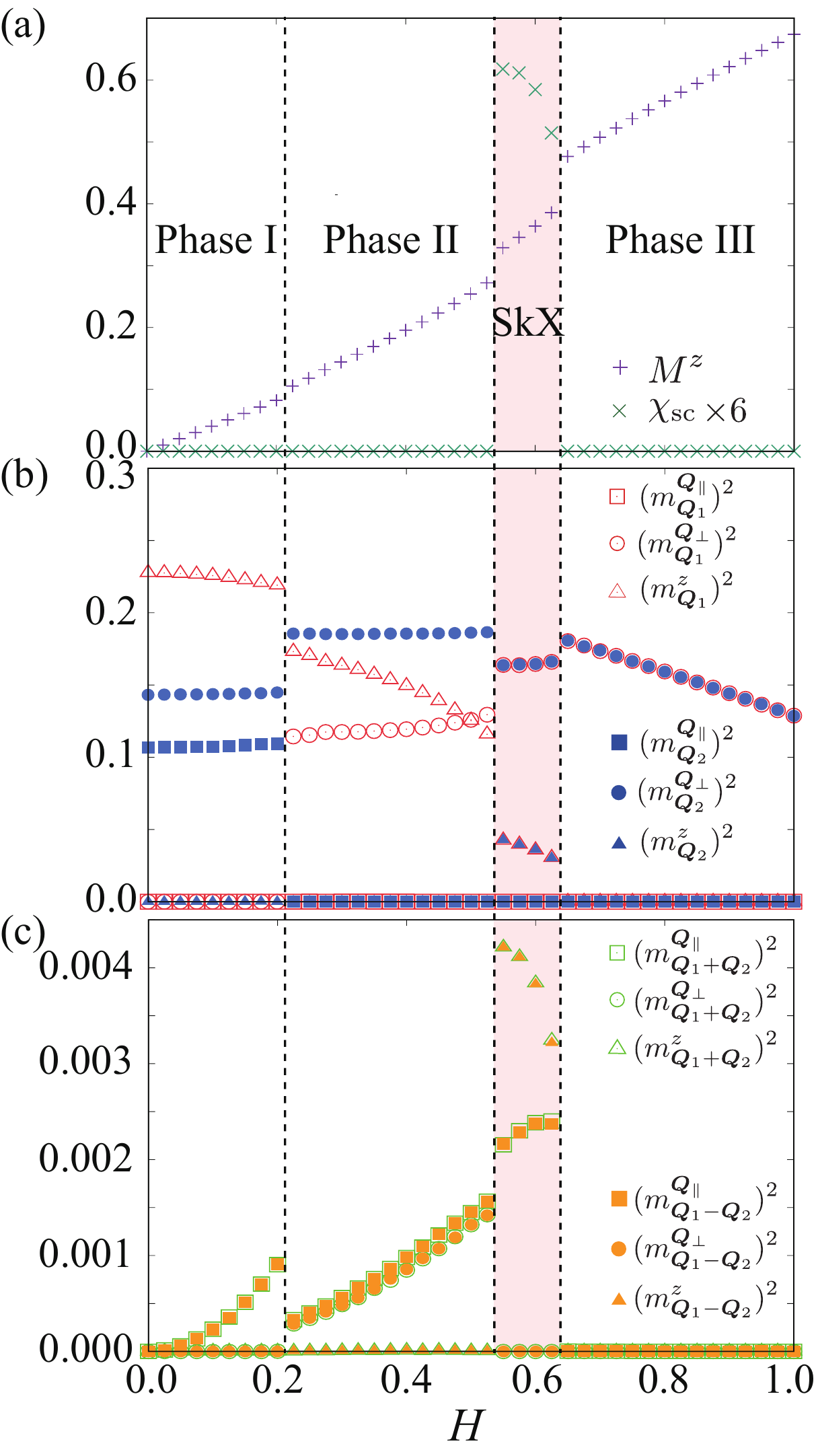} 
\caption{
\label{Fig:mq}
(Color online) Magnetic field ($H$) dependence of (a) $M^z$ and $\chi_{\rm sc}$ and (b), (c) $(m^{\bm{Q}_\parallel}_{\zeta})^2$, $(m^{\bm{Q}_\perp}_{\zeta})^2$, and $(m^{z}_{\zeta})^2$ for (b) $\zeta=\bm{Q}_1$ and $\bm{Q}_2$ and (c) $\zeta=\bm{Q}_1+\bm{Q}_2$ and $\bm{Q}_1-\bm{Q}_2$ at $K=0.5$, $\Gamma^{\rm A}_{\bm{Q}_1}=0.045$ ($\Gamma^{xx}=0.855$ and $\Gamma^{yy}=0.9$), and $\Gamma^{\rm A}_{\bm{Q}_1+\bm{Q}_2}=-0.01$. 
}
\end{center}
\end{figure}
Next, we perform the simulated annealing by means of Monte Carlo simulations for $N=96^2$ in order to demonstrate the emergence of four types of the SkXs. 
The simulations have been done by following the manner in Ref.~\citen{Hayami_PhysRevB.95.224424}. 
We set the final temperature as $T=0.01$, and confirmed that the qualitatively same results were obtained for a lower temperature.
In addition to the model parameters, $J=1$, $K=0.5$, $\Gamma^{xx}=0.855$, $\Gamma^{yy}=0.9$, and $\Gamma^{zz}=1$, which stabilize the Bloch SkX or anti-SkX II in the intermediate $H$ region, we introduce the contributions of the higher harmonics in Eq.~(\ref{eq:q1plusq2}). 
Considering the situation where the magnetic bare susceptibility shows sharp peaks at $\bm{Q}_1$ and $\bm{Q}_2$, we assume the small contributions at $\bm{Q}_1+\bm{Q}_2$ and $\bm{Q}_1-\bm{Q}_2$, $\Gamma'^{xx}=\Gamma'^{yy}=\Gamma'^{zz}=0$ and $\Gamma^{\rm A}_{\bm{Q}_1+\bm{Q}_2}=-0.01$, for simplicity.

Figures~\ref{Fig:mq}(a)-\ref{Fig:mq}(c) show the results against $H$.  
The net magnetization $M^z=(1/N)\sum_i S_i^z$ and the scalar chirality $\chi_{\rm sc}=[(1/N)\sum_{i,\delta=\pm1}\bm{S}_i \cdot (\bm{S}_{i+\delta \hat{x}}\times \bm{S}_{j+\delta \hat{y}})]^2$ [$\hat{x}$ ($\hat{y}$) is the unit vector in the $x$ ($y$) direction] are shown in Fig.~\ref{Fig:mq}(a), while the $\bm{Q}_1$ and $\bm{Q}_2$ ($\bm{Q}_1+\bm{Q}_2$ and $\bm{Q}_1-\bm{Q}_2$) components of the magnetic moment 
$(m^{\bm{Q}_\parallel}_{\bm{q}}, m^{\bm{Q}_\perp}_{\bm{q}},m^{z}_{\bm{q}})$, which are orthogonal with each other, are shown in Fig.~\ref{Fig:mq}(b) [Fig.~\ref{Fig:mq}(c)]. 
Here, $m^{z}_{\bm{q}}=\sqrt{S^{zz}(\bm{q})/N}$ represents the $z$ spin component, where $S^{\alpha\alpha}(\bm{q})=(1/N)\sum_{i,j}S^{\alpha}_i S^{\alpha}_j e^{i\bm{q}\cdot (\bm{r}_i -\bm{r}_j)}$ is the spin structure factor for the $\alpha=x,y,z$ spin moments, whereas $m^{\bm{Q}_\parallel}_{\bm{q}}$ ($m^{\bm{Q}_\perp}_{\bm{q}}$) represents the inplane components parallel (perpendicular) to the $\bm{q}$ direction.

\begin{figure}[t!]
\begin{center}
\includegraphics[width=1.0 \hsize]{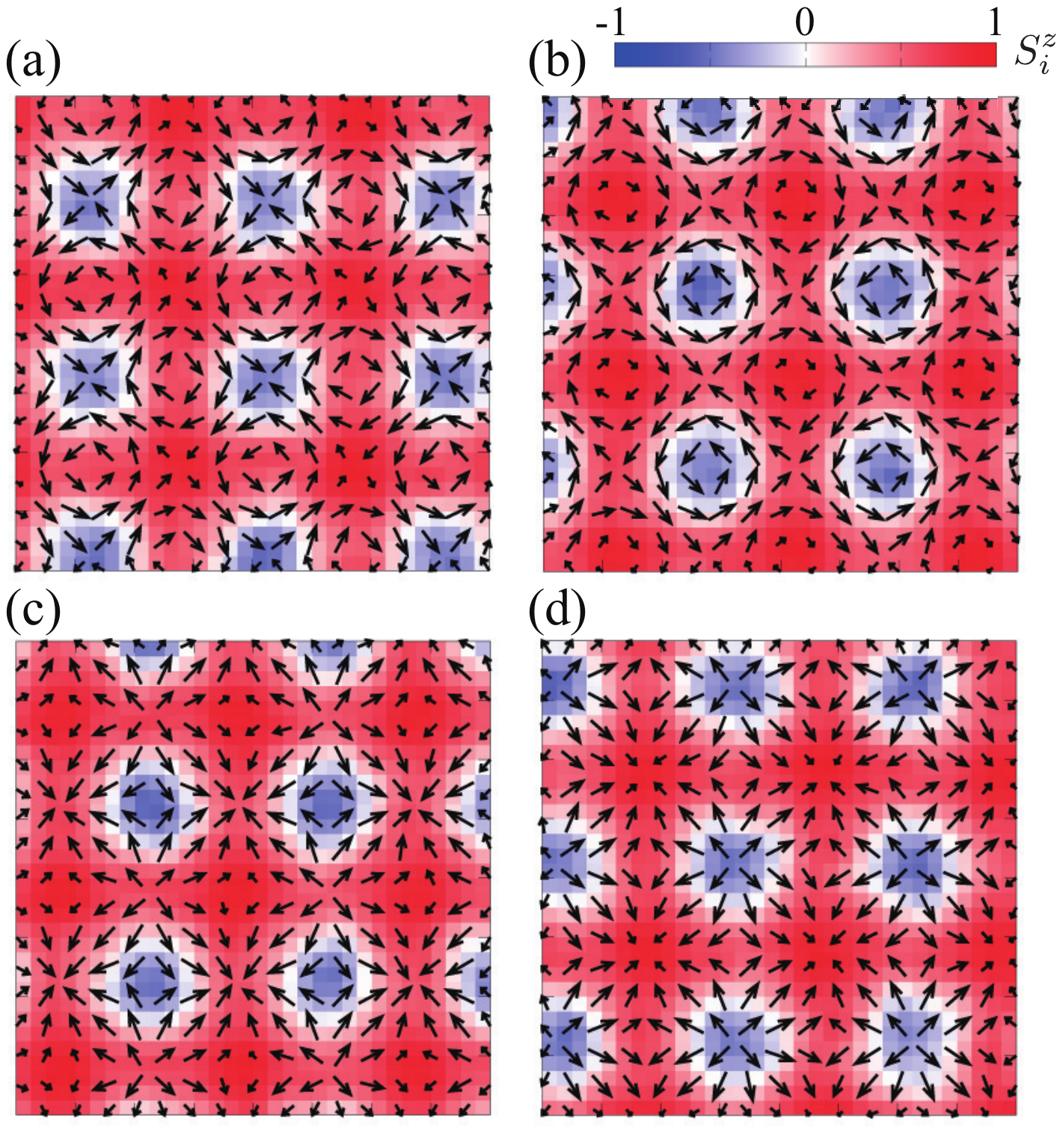} 
\caption{
\label{Fig:spin}
(Color online) Real-space spin configurations in (a) the anti-SkX II for $\Gamma^{\rm A}_{\bm{Q}_1}=0.045$ and $\Gamma^{\rm A}_{\bm{Q}_1+\bm{Q}_2}=-0.01$, (b) the Bloch SkX for $\Gamma^{\rm A}_{\bm{Q}_1}=0.045$ and $\Gamma^{\rm A}_{\bm{Q}_1+\bm{Q}_2}=0.01$, (c) the anti-SkX I for $\Gamma^{\rm A}_{\bm{Q}_1}=-0.045$ and $\Gamma^{\rm A}_{\bm{Q}_1+\bm{Q}_2}=0.01$, and (d) the N\'eel SkX for $\Gamma^{\rm A}_{\bm{Q}_1}=-0.045$ and $\Gamma^{\rm A}_{\bm{Q}_1+\bm{Q}_2}=-0.01$ obtained by the simulated annealing with $N = 96^2$. 
The contour shows the $z$ component of the spin moment, and the arrows represent the $xy$ spin components.
The other parameter is $H=0.6$.
In each spin texture, a part of the whole lattice is shown. 
}
\end{center}
\end{figure}

There are four magnetic phases with nonzero magnitudes at $\bm{Q}_1$ and $\bm{Q}_2$, i.e., the double-$Q$ states, as shown in Fig.~\ref{Fig:mq}(b). 
Among them, the intermediate state stabilized for $0.55 \lesssim H\lesssim 0.63$ corresponds to the SkX exhibiting the net scalar chirality $\chi_{\rm sc}$, as shown in Fig.~\ref{Fig:mq}(a). 
The important point is that the SkX always has the small contributions of $m^{\bm{Q}_\parallel}_{\bm{q}}$ or $m^{\bm{Q}_\perp}_{\bm{q}}$ at $\bm{Q}_1+\bm{Q}_2$ and $\bm{Q}_1-\bm{Q}_2$ in Fig.~\ref{Fig:mq}(c), in which their difference results in the degeneracy lifting between the Bloch SkX and anti-SkX II by $\Gamma_{\bm{Q}_1+\bm{Q}_2}^{\rm A}$. 
The real-space spin configuration obtained by the simulated annealing is shown in Fig.~\ref{Fig:spin}(a) where the skyrmion core is located at the center of the square plaquette in the lattice structure. 
As expected from the above argument, the anti-SkX II is obtained in the present model parameters for $\Gamma_{\bm{Q}_1}^{\rm A}> 0$ and $\Gamma_{\bm{Q}_1+\bm{Q}_2}^{\rm A}< 0$.

The other three SkXs can be also obtained by reversing the sign of $\Gamma_{\bm{Q}_1}^{\rm A}$ and/or $\Gamma_{\bm{Q}_1+\bm{Q}_2}^{\rm A}$. 
We show the obtained SkX spin textures for different sets of signs at fixed $H=0.6$ in Figs.~\ref{Fig:spin}(b)--\ref{Fig:spin}(d); 
the Bloch SkX for $\Gamma_{\bm{Q}_1}^{\rm A}> 0$ and $\Gamma_{\bm{Q}_1+\bm{Q}_2}^{\rm A}> 0$ in Fig.~\ref{Fig:spin}(b), the anti-SkX I for $\Gamma_{\bm{Q}_1}^{\rm A}< 0$ and $\Gamma_{\bm{Q}_1+\bm{Q}_2}^{\rm A}> 0$ in Fig.~\ref{Fig:spin}(c), and the N\'eel SkX for $\Gamma_{\bm{Q}_1}^{\rm A}<0$ and $\Gamma_{\bm{Q}_1+\bm{Q}_2}^{\rm A}< 0$ in Fig.~\ref{Fig:spin}(d). 
Thus, the signs of the effective anisotropic RKKY interactions not only at the ordering vectors $\bm{Q}_1$ and $\bm{Q}_2$ but also at the higher harmonic vectors $\bm{Q}_1+\bm{Q}_2$ and $\bm{Q}_1-\bm{Q}_2$ determine the types of the SkXs in centrosymmetric magnets. 

Finally, we discuss the origin of the anisotropic form factors $\Gamma_{\bm{Q}_1}^{\rm A}$ and $\Gamma_{\bm{Q}_1+\bm{Q}_2}^{\rm A}$. 
From the fact that the effective spin model is obtained by the perturbation expansion with respect to the spin-charge (Kondo) coupling in the Kondo lattice model, and the spin-charge coupling is also obtained by the perturbation expansion with respect to the hybridization between itinerant and localized electrons in the periodic Anderson model, the nature of the anisotropic form factors arises from the hybridization between itinerant and localized electrons. 
In particular, the spin-orbit coupling in multi-orbital or multi-sublattice systems leads to effective spin-dependent hybridization between itinerant and localized electrons even on a centrosymmetric lattice, which gives rise to the microscopic origin of the anisotropic form factors. 
As a result, the magnitudes of $\Gamma_{\bm{Q}_1}^{\rm A}$ and $\Gamma_{\bm{Q}_1+\bm{Q}_2}^{\rm A}$ depend on the orbital nature and the sublattice position as well as the electronic band structures. 
Furthermore, we confirmed that the signs of $\Gamma_{\bm{Q}_1}^{\rm A}$ and $\Gamma_{\bm{Q}_1+\bm{Q}_2}^{\rm A}$ are sensitive to above factors, which indicates that the types of the SkXs in materials can be changed by an external pressure and electron doping. 
The detailed analysis of the anisotropic form factor will be reported elsewhere.

In summary, we have proposed the degeneracy lifting of the vorticity and helicity degrees of freedom in the SkXs in itinerant centrosymmetric magnets. 
We have demonstrated that four types of the SkXs can be realized by considering the signs of the two symmetric anisotropic interactions in the effective spin model without relying on the DM-type antisymmetric interactions rooted in the specific noncentrosymmetric point groups. 
We showed that the contributions from the higher harmonics cannot be neglected to determine the types of the SkXs. 
As the magnitude and sign are dependent on the atomic orbitals and electronic band structures, it would be possible of controlling and engineering the SkXs in centrosymmetric magnets by changing the external parameters, which is in contrast to the conventional SkXs in noncentrosymmetric magnets.  

\begin{acknowledgments}
S.H. is grateful to Y. Motome, S. Seki, and N. D. Khanh for fruitful discussions. 
This research was supported by JSPS KAKENHI Grant Numbers JP18K13488, JP19K03752, and JP19H01834. 
Parts of the numerical calculations were performed in the supercomputing systems in ISSP, the University of Tokyo.
\end{acknowledgments}

\bibliographystyle{JPSJ}
\bibliography{ref}

\begin{thebibliography}{10}
\bibitem{comment_Gammazz}
We implicitly assume that $\Gamma^{zz}$ is at least larger than either
  $\Gamma^{xx}$ or $\Gamma^{yy}$ so that the spiral plane lies in the $yz$ or
  $xz$ plane.
\end{thebibliography}

\end{document}